\documentclass[aps,prl,showpacs,superscriptaddress,twocolumn]{revtex4-1}
\usepackage{amsmath}
\usepackage{amssymb}
\usepackage{graphicx}
\usepackage{bm}

\newcommand{\expv}[1]{\langle #1 \rangle}	
\newcommand{\ket}[1]{| #1 \rangle} 
\newcommand{\bra}[1]{\langle #1 |} 
\newcommand{\braket}[2]{\langle #1 \vphantom{#2} | #2 \vphantom{#1} \rangle} 
\newcommand{\op}[1]{\hat #1}			%

\begin{document}
\title{Generalized Thermalization in an Integrable Lattice System}

\author{Amy C. Cassidy}
\author{Charles W. Clark}
\affiliation{Joint Quantum Institute, National Institute of Standards and Technology, Gaithersburg, MD 20899, USA}
\author{Marcos Rigol}
\affiliation{Department of Physics, Georgetown University, Washington, DC 20057, USA}
 
\date{\today}

\begin{abstract}
After a quench, observables in an integrable system may not relax to the standard thermal values, but can 
relax to the ones predicted by the Generalized Gibbs Ensemble (GGE)  [M. Rigol \textit{et al.}, 
PRL \textbf{98}, 050405 (2007)]. The GGE has been shown to accurately describe observables in various 
one-dimensional integrable systems, but the origin of its success is not fully understood.  Here we 
introduce a microcanonical version of the GGE and provide a justification of the GGE based on a 
generalized interpretation of the eigenstate thermalization hypothesis, which was previously introduced 
to explain thermalization of nonintegrable systems.  We study relaxation after a quench of one-dimensional 
hard-core bosons in an optical lattice.  Exact numerical calculations for up to 10 particles on 50 lattice 
sites ($\approx 10^{10}$ eigenstates) validate our approach.

\end{abstract}

\maketitle
Once only of theoretical interest, integrable models of one-dimensional (1D) quantum many-body systems
can now be realized with ultracold atoms \cite{paredes04,*kinoshita04}. The possibility of controlling the 
effective dimensionality and the degree of isolation have allowed access to the quasi-1D regime and to the 
long coherence times necessary to realize integrable models. Additionally, advances in the cooling and trapping 
of atoms have led to increased interest in dynamics following quantum quenches, where a many-body system in 
equilibrium is exposed to rapid changes in the confining potential or interparticle interactions.
  
In general, in integrable quantum systems that are far from equilibrium, observables cannot relax to the 
usual thermal state predictions because they are constrained by the non-trivial set of conserved quantities that 
make the system integrable \cite{sutherland04}. Relaxation to non-thermal values were recently observed in a 
cold-atom system close to integrability \cite{kinoshita06}. At integrability, it is natural to describe the 
observables after relaxation by an updated statistical mechanical ensemble: the generalized Gibbs ensemble (GGE) 
\cite{rigol07STATa}, which is constructed by maximizing the entropy subject to the integrability constraints 
\cite{jaynes57a,*jaynes57b}. In recent studies of integrable systems
\cite{rigol07STATa,cazalilla06,*calabrese07a,*cramer08a,*barthel08,*fioretto2010quantum,*rigol06STATb,kollar08},
the GGE has been found to accurately describe various observables after relaxation, 
but a microscopic understanding of its origin and applicability remains elusive. In particular, an important 
question remains: how is it that expectation values after relaxation can be described by an ensemble with exponentially 
fewer parameters than the size of the Hilbert space? The full dynamics are determined by as many parameters as the 
size of the latter. At a microscopic level, thermalization for non-integrable systems can be understood in terms 
of the eigenstate thermalization hypothesis (ETH) \cite{deutsch91,*srednicki94,rigol08STATc}, which, however, 
breaks down as one approaches integrability \cite{rigol09STATa,*rigol09STATb}. 

This paper is devoted to the study of how generalized thermalization, in the sense of relaxation to the predictions
of the GGE, takes place in integrable systems. Answering this question is important not merely because of its
relevance to the foundations of statistical mechanics in integrable systems, but also because it has become
necessary to understand recent experiments with ultracold gases in quasi-1D geometries. 
For integrable systems, we compare the predictions of quantum mechanics with those of various statistical ensembles. 
In particular, we introduce a microcanonical version of the GGE, which we use to show that relaxation to the GGE 
can be understood in terms of a generalized view of the ETH.

We study the dynamics following an instantaneous quench of 1D hard-core bosons on a lattice, which is fully integrable. 
The Hamiltonian is given by
\begin{equation}
 \hat{H} = -J \sum_{i=1}^{L-1} \left( \hat{b}_i^\dagger \hat{b}_{i+1} + \textrm{H.c.}\right) 
     + V(\tau)\sum_{i=1}^L (i-L/2)^2 \op{n}_i
\label{eq:hamil}
\end{equation}
where $J$ is the hopping parameter;  $V(\tau)$ gives the curvature of an additional parabolic trapping potential for 
atoms on a lattice with lattice constant $a$; $\hat{b}_i^\dagger\ (\hat{b}_i)$ is the hard-core bosonic creation 
(annihilation) operator; and $\op{n}_i = \hat{b}_i^\dagger \hat{b}_i$ is the number operator. In addition to the 
standard commutation relations for bosons, hard-core bosons satisfy the constraint $\op{b}_i^{\dagger2}=\hat{b}_i^2=0$,
which forbids multiple occupancy of the lattice sites. This Hamiltonian can be mapped onto non-interacting fermions 
through the Jordan-Wigner  transformation \cite{jordan28}, and the many-body eigenstates can be constructed as 
Slater determinants  of the single-particle fermionic eigenstates \cite{rigol05HCBc}.

We will focus on the behavior of the momentum distribution function,
$ \expv{\op{n}_k} = \sum_{l,m} e^{-i k(l-m)} \bra{\psi}\op{b}_m^\dagger \op{b}_l\ket{\psi}/L$, for
system sizes ranging from $N=5$ bosons on $L=25$ lattice sites to $N=10$ bosons on $L=50$ lattice sites 
($\approx 10^{10}$ eigenstates). Initially, we prepare the system in the ground state $\ket{\psi_0}$ of a 1D 
lattice with hard-wall boundary conditions and an additional harmonic potential, with trapping strength $V=V_0$. 
At time $\tau=0$, the harmonic trap is turned off, $V(\tau\geq0)=0$, and the state $\ket{\psi(\tau)}$ evolves under 
the influence of the final Hamiltonian. Hereafter, we refer to this state as it is immediately after the quench as 
the ``quenched state".  Its time evolution is given by
$
\ket{\psi(\tau)} = \sum_{\alpha} c_{\alpha}e^{-iE_\alpha \tau/\hbar} \ket{\alpha},
$
where $\ket{\alpha}$ are the energy eigenstates of the final Hamiltonian with energies $E_{\alpha}$, and 
$c_{\alpha} = \braket{\alpha}{\psi_0}$ are the overlaps between the eigenstates of the final Hamiltonian and 
the quenched state. After relaxation, assuming the degeneracies in energy levels are irrelevant,
the expectation value of an observable is expected to be given by the so called 
diagonal ensemble (DE) \cite{rigol08STATc,kollar08,rigol09STATa,*rigol09STATb}
\begin{equation*}
\expv{\hat{A}}_\text{DE} =\lim_{\tau \rightarrow \infty} \frac{1}{\tau}\int_0^\tau d\tau^\prime 
\bra{\psi(\tau^\prime)}\hat{A}\ket{\psi(\tau^\prime)}= 
\sum_\alpha |c_\alpha|^2 \bra{\alpha} \hat{A} \ket{\alpha}.
\end{equation*}
We have checked numerically that, despite the integrability of our model, $n_k$ relaxes 
to the DE prediction, with small fluctuations around this result \cite{supp_mat}.

\begin{figure}
\centering
\includegraphics[scale=0.65]{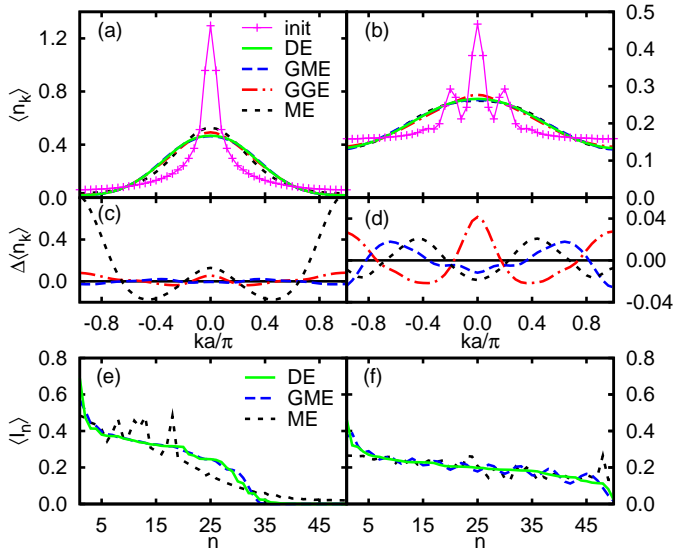}\vspace{-1em}
\caption{\label{fig:mom_dist}(a),(b) Momentum distribution of the initial state (init), diagonal (DE), 
generalized microcanonical (GME), generalized Gibbs (GGE), and the microcanonical (ME) ensembles.
(c),(d) Relative difference of the GME, GGE and ME from the DE. (e),(f) Conserved 
quantities, $\expv{\hat{I}_{n}}$, in the quenched state (identical to the DE), GME and ME.  
$\expv{I_n}$ are ordered in descending occupations in the quenched state. 
$L=50$, $N=10$, $\delta_\text{ME}=0.05J$, $ \delta_\text{GME}=0.8$.  (a),(c),(e) 
$\varepsilon=0.72J$, $V_0=0.029J$. (b),(d),(f) $\varepsilon=1.52J$, $V_0=0.125J$.}\vspace{-1em}
\end{figure}

Figure 1 shows the momentum distributions, $n_k$, before and after the quench, for two different initial 
trap strengths, which correspond to different energies per particle, $\varepsilon$, after the quench. 
These results are compared with those of various ensembles of statistical mechanics. The microcanonical 
ensemble (ME) is one in which all eigenstates in the relevant energy window
have identical weights.  Within the microcanonical ensemble, the expectation value of a generic 
observable $A$ is $\expv{\hat{A}}_\text{ME}=N^{-1}_{\varepsilon,\delta_\text{ME}}
\sum_{\alpha,|\varepsilon-\varepsilon_\alpha|<\delta_\text{ME}}\bra{\alpha}\hat{A}\ket{\alpha}$, 
where $\delta_\text{ME}$ is small, but still much greater than the mean many-body level spacing. 
$N_{\varepsilon,\delta_\text{ME}}$ is the number of eigenstates in the energy window 
$|\varepsilon-\varepsilon_\alpha|<\delta_\text{ME}$. We have checked that the results reported here are 
nearly independent of the specific value of $\delta_\text{ME}$. The GGE is a grand-canonical ensemble 
that maximizes the entropy subject to the constraints associated with non-trivial conserved quantities 
of the quenched state. The density matrix takes the form \cite{rigol07STATa}
\begin{equation}
\hat{\rho}_\text{GGE} = Z_G^{-1} e^{-\sum\lambda_n \hat{I}_n} , 
\qquad Z_G= \text{Tr} \left[ e^{-\sum\lambda_n \hat{I}_n} \right],
\end{equation}
where \{$\hat{I}_n$\}, $n=1,\ldots,L$, are the conserved quantities. In our systems, these correspond to the 
occupation of the single-particle eigenstates of the underlying noninteracting fermions to which hard-core 
bosons are mapped, and \{$\lambda_n$\} are Lagrange multipliers fixed by the initial conditions, 
$\lambda_n=\ln [(1-\bra{\psi_0}\hat{I}_{n}\ket{\psi_0})/\bra{\psi_0}\hat{I}_{n}\ket{\psi_0}]$ \cite{rigol07STATa}. 
Observables within this ensemble are then computed as $\expv{\hat{A}}_\text{GGE}=\text{Tr}\left[\hat{A}\,
\hat{\rho}_\text{GGE}\right]$ following Ref.~\cite{rigol05HCBc}.

As a step towards understanding the GGE as well as developing a more accurate description of isolated integrable 
systems after relaxation, we introduce a microcanonical version of the GGE, which we call the generalized microcanonical 
ensemble (GME).  Like the ME, where states within a small energy window contribute with equal weight, within the 
GME we assign equal weight to all eigenstates whose values of the conserved quantities are close to the desired values. 
The expectation value of a generic observable within the generalized microcanonical ensemble is given by 
$\expv{\hat{A}}_\text{GME}=\mathcal{N}^{-1}_{\{I_n\},\delta_\text{G ME}} 
\sum_{\alpha, \delta_\alpha < \delta_\text{GME}} \bra{\alpha}\hat{A} \ket{\alpha}$,  where 
$ \sum_{\alpha, \delta_\alpha < \delta_\text{GME}}$ is a sum over eigenstates that are within the GME window 
and $\mathcal{N}_{\{I_n\},\delta_\text{GME}}$ is the number of states within that window and $\delta_\alpha$ 
is a measure of the distance of eigenstate $\alpha$ from the target distribution.

In order to construct the GME, we include eigenstates of the Hamiltonian with a similar distribution of conserved 
quantities which once averaged reproduce the values of the conserved quantities in the quenched state. This approach 
is characterized by three ingredients: (i) The ordered distribution (from largest to smallest) of the conserved 
quantities in the DE, $\expv{I_n}_\textrm{DE}\equiv\sum_\alpha |c_\alpha|^2 I_{n,\alpha} $ [as in Figs.~1(e) and 1(f)], 
(ii) a target distribution of the nonzero expectation values of the conserved quantities $\{I^*_{n^*_i}=1\}$, 
where the values of $n^*_i$ ($i=1,\ldots,N$) are chose to describe the distribution $I_n$ in a coarse grained sense 
\cite{nstar}, and (iii) for each individual many-body eigenstate, the distance from the target state, $\delta_\alpha$, 
which we define as $\delta_\alpha=\left[\frac{1}{N}\sum_{i=1}^N I_{n^*_i}(n_{i,\alpha} - n^*_i)^2 \right] ^{1/2}$.
Here $n_{i,\alpha}$ ($i=1,\ldots,N$) are the single-particle states occupied in eigenstate $\alpha$, and $I_{n^*_i}$ 
are the interpolated values of $\expv{I_n}_\textrm{DE}$, evaluated at $n^*_i$. The definition of $\delta_\alpha$ is not 
unique and several variants that do not change our conclusions were also considered \cite{supp_mat}. 

To better visualize the differences between the results of the various ensembles in Figs.~1(a) and 1(b), 
we have plotted $\Delta\expv{n_k}_\text{stat}=(\expv{\hat{n}_k}_\text{DE}-\expv{\hat{n}_k}_\text{stat})/
\expv{\hat{n}_k}_\text{DE}$, where ``stat'' stands for ME, GGE, or GME in Figs.~1(c) and 1(d). For weaker initial 
confinements (smaller $\varepsilon$ - Fig.~1(c)), the GME is practically indistinguishable from the diagonal 
distribution. Both the GME and the GGE accurately capture the tails of $n_k$, while the thermal ensemble does not. 
For tighter initial traps (greater $\varepsilon$ - Fig.~1(d)) all four ensembles are very similar (note the scale), 
suggesting that $n_k$ in the final steady state is indistinguishable from that of the thermal state.

The close agreement between DE and ME results in Fig.~1(b) raises the question: how can an integrable system thermalize, 
given the constraints imposed by the complete set of conserved quantities? We conjecture that if the values of the 
conserved quantities in the quenched state are similar to those of the ME, then the latter will accurately describe 
observables after relaxation. This may occur for a variety of quenches.

In Figs.~1(e) and 1(f), we plot the values of the conserved quantities in the quenched state and compare them with 
the expectation values of those quantities in different statistical ensembles. (By definition, the distribution of 
conserved quantities in the DE and GGE are identical to that of the quenched state.) Figure 1(e) shows that the 
microcanonical values of the conserved quantities are clearly different from the values in the quenched state, while 
in Fig.~1(f) they are very similar. This supports our conjecture above, and demonstrates that thermalization can occur 
in integrable systems for special initial conditions. Additionally, the GME reproduces the correct distribution of 
the conserved quantities supporting the validity of our method for generating it.

To quantify the above observations, and to understand what happens in the thermodynamic limit, we have studied the 
difference between the predictions of the DE and the statistical ensembles for different system sizes.  We compute the 
integrated relative differences, $(\Delta n_k)_\text{stat}=\sum_k|\expv{\hat{n}_k}_\text{DE} - 
\expv{\hat{n}_k}_\text{stat}|/\sum_k\expv{\hat{n}_k}_\text{DE}$, where again ``stat'' stands for ME, GGE, or GME.

\begin{figure}
\centering
\includegraphics[scale=0.7]{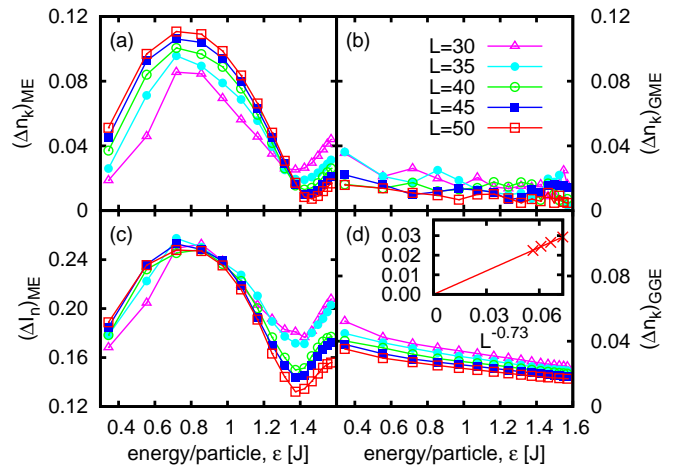}\vspace{-1em}
\caption{\label{fig:delta_nk} (a) $(\Delta n_k)_\text{ME}$ versus energy per particle of the quenched state. 
$\delta_\text{ME}=0.05J$. (b) $(\Delta n_k)_\text{GME}$ vs $\varepsilon$, $ \delta_\text{GME}=0.8$. 
(c) Integrated difference between the conserved quantities in the quenched state and the ME, $(\Delta I_{n})_\text{ME}$. 
(d) $(\Delta n_k)_\text{GGE}$ vs $\varepsilon$. Inset: $(\Delta n_k)_\text{GGE}$ vs $L^{-0.73}$ for $\varepsilon=1.07J$, 
where a fit to $(\Delta n_k)_\text{GGE}=zL^{-\gamma}$ gives $\gamma=0.73 \pm 0.02$.
}\vspace{-1em}
\end{figure}

In Fig.~2(a), we plot $(\Delta n_k)_\text{ME}$ as a function of the final energy per particle, $\varepsilon$, for 
different lattice sizes, $L$. To perform finite-size scaling, $\varepsilon$ and the filling factor ($\nu=N/L=0.2$) are 
held constant as $L$ changes. Figure 2(a) shows that for $\varepsilon \lesssim 1.3J$ the difference between the $n_k$ 
in the DE and the ME increases with increasing $L$, indicating that the difference persists in the thermodynamic limit. 
For $\varepsilon \gtrsim 1.3J$, the opposite behavior is observed. From our previous discussion, one expects that
$(\Delta n_k)_\text{ME}$ should closely follow the behavior of the integrated differences between the conserved 
quantities in the quenched state and the ME,
$(\Delta I_n)_\text{ME} = {\sum_n |\expv{I_{n}}_\text{DE} - \expv{I_{n}}_\text{ME}|} /{\sum_n \expv{I_{n}}_\text{DE}}$.
This is seen by comparing Figs.~2(a) and 2(c), which leads us to conclude that $n_k$ need not relax to the standard 
thermal prediction, except when $(\Delta I_n)_\text{ME}$ becomes negligible. Qualitatively similar results were obtained 
in the canonical ensemble \cite{supp_mat}.

On the other hand, in Fig.~2(b) one can see that the differences between $n_k$ in the diagonal and generalized 
microcanonical ensembles are very small and decrease with increasing system size, so that the former successfully 
describes this observable after relaxation. In the case of the GGE [Fig.~2(d)], $(\Delta n_k)_\text{GGE}$ is in general 
larger than $(\Delta n_k)_\text{GME}$, which is to be expected since the GGE is a grand-canonical  ensemble. As the 
system size increases $(\Delta n_k)_\text{GGE}\rightarrow 0$ as $L^{-\gamma}$, where $\gamma \approx 0.73$  
[inset of Fig.~2(d)] and slightly depends on the energy \cite{supp_mat}.

The question that remains to be answered is why the generalized Gibbs and the generalized 
microcanonical ensemble are able to describe the $n_k$ after relaxation, i.e., why
$\expv{\hat{n}_k}_\text{GGE}=\expv{\hat{n}_k}_\text{GME} = \expv{\hat{n}_k}_\text{DE} 
\equiv \sum_\alpha |c_\alpha|^2 \bra{\alpha}\hat{n}_k\ket{\alpha}$. Note that whereas $\expv{\hat{n}_k}_\text{GGE}$ 
and $\expv{\hat{n}_k}_\text{GME}$ are entirely determined by the $L$ independent values of the conserved quantities 
in the quenched state, $\expv{\hat{n}_k}_\text{DE}$ is determined by the exponentially larger $\binom{L}{N}$ values 
of the coefficients $c_\alpha$. 

To address this question, we perform a spectral decomposition of $\expv{\hat{n}_k}_\text{DE} $ and 
$\expv{\hat{n}_k}_\text{GME}$. Figure 3 displays a coarse grained view of the weight which eigenstates with a 
given zero momentum occupancy $\expv{\hat{n}_{k=0}}_\alpha=\bra{\alpha}\hat{n}_{k=0}\ket{\alpha}$ contribute 
to the DE [Fig.~3(a)] and the GME [Fig.~3(b)]. The correlation between the results in both figures is apparent. 
However, it is not clear why the details contained in the overlaps $c_\alpha$ are completely washed out so that the
DE and the GME results coincide, while they are different from those in the ME. In the inset of 
Fig.~3(a), we plot a histogram of the values of $n_{k=0}$ for the DE, GME and the ME. Clearly the histograms 
for the DE and GME have a similar mean but different widths, while the ME has a different mean and width \cite{supp_mat}.

\begin{figure}
\centering
\includegraphics[scale=0.65]{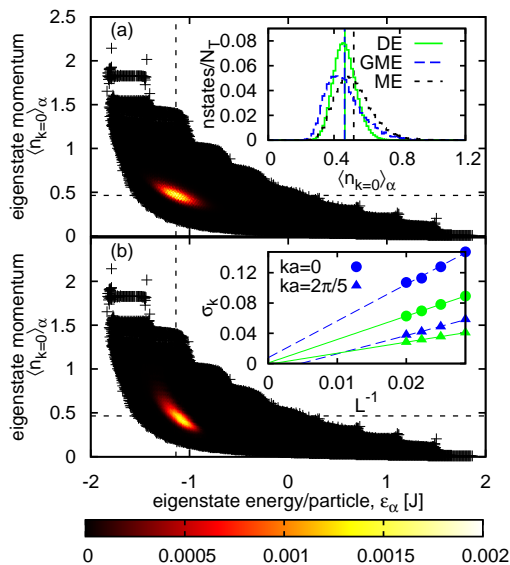}\vspace{-1em}
\caption{\label{fig:eev_dist} Density plot of the coarse-grained weights with which eigenstates contribute to 
(a) the DE (sum over diagonal weights, $|c_\alpha|^2)$ and (b) the GME (fractional number of states = number of 
states/total number of states) as a function of eigenstate energy and $\expv{\hat{n}_{k=0}}_\alpha$. The sums are 
performed over window of width $\delta n_k= 0.0067$ , and $\delta \varepsilon= 0.035J$. The horizontal and vertical 
dotted lines are the expectation values of $\hat{n}_{k=0}$ and $\varepsilon$ in each ensemble.  
$L=45$, $N=9$, $V_0 = 0.036J$, $\varepsilon= 0.72J$,  $\delta_\text{GME}=0.85$. Inset in (a): Histogram of DE 
weights (green), fractional number of GME states (blue) and fractional number of ME states (black) summed over 
all energies.  Bin width, $\delta n_k= 0.0067$. Vertical lines give the mean,  $\expv{\hat{n}_{k=0}}$ within each 
ensemble. Inset in (b):  Fluctuations of $\expv{\hat{n}_{ka=0}} (\bullet)$ and 
$\expv{\hat{n}_{ka=2\pi/5}} (\blacktriangle)$ within the DE (green) and GME (blue) as a function of inverse system 
size. $\varepsilon= 0.72J$.}\vspace{-1em}
\end{figure}

Ultimately, one is interested in what happens in the thermodynamic limit. For each $k$, we define the width of the 
distribution of $\expv{\hat{n}_k}_\alpha$ for each ensemble as 
$\sigma_{k}=\sqrt{\expv{\hat{n}_k^2} - \expv{\hat{n}_k}^2}$. The inset of Fig.~3(b), shows $\sigma_{k}$ within the DE 
and the GME versus $L^{-1}$. The scaling is depicted for two $k$ values and clearly shows that the widths of both 
distributions vanish in the thermodynamic limit. This demonstrates that the overwhelming majority of the states 
selected by the DE as well as by the GME, which have similar values of the conserved quantities, have identical 
expectation values of $n_k$. This is why details of the distribution of $c_\alpha$ no longer matter as $L$ increases. 
We note that with increasing $L$, the number of eigenstates contained in the generalized microcanonical window 
increases exponentially, however, the ratio of the number of states in the GME and the ME vanishes \cite{supp_mat}.

The findings above provide a generalization of the ETH introduced previously to understand thermalization in 
nonintegrable systems \cite{deutsch91,*srednicki94,rigol08STATc}. The ETH states that the expectation values of 
few-body observables in generic systems do not fluctuate between eigenstates that are close in energy. Thus all 
eigenstates within a microcanonical window have essentially the same expectation values of the observables, and one 
can say that thermalization occurs at the level of eigenstates.  As seen in Fig.~3, $\expv{\hat{n}_{k}}_{\alpha}$ 
exhibits large eigenstate-to-eigenstate fluctuations in our integrable system, showing that ETH is invalid. However, 
by selecting eigenstates with similar conserved quantities, ETH is restored, although in a weaker sense: the 
overwhelming majority of eigenstates with similar conserved quantities have similar values of $n_k$. These results 
pave the way to a unified understanding of thermalization in generic (nonintegrable systems) and its generalization
in integrable systems.  This opens many new questions, such as whether the concepts of typicality 
\cite{tasaki98,*goldstein06,*popescu06,*reimann08} and thermodynamics \cite{polkovnikov08microscopic,*polkovnikov08} 
can be generalized to isolated integrable systems.

\begin{acknowledgments}
This work was supported by NSF under Physics Frontier Grant No. PHY-0822671.
 A.C.C. acknowledges support from NRC/NIST and M.R. acknowledges support from the Office of Naval Research.  
We thank V. Dunjko, L. Mathey, and M. Olshanii for helpful discussions. 
\end{acknowledgments}

%


%
%
\pagebreak
\onecolumngrid
\setcounter{figure}{0}
 \setcounter{equation}{0}
 
\begin{center}
{\large \bf Supplementary material for EPAPS
\\ Generalized Thermalization in an Integrable Lattice System}\\
\vspace{0.4cm}

Amy C. Cassidy,${}^1$ Charles W. Clark,${}^1$ Marcos Rigol${}^2$

{\small \it ${}^1$Joint Quantum Institute, National Institute of Standards and Technology, Gaithersburg, MD 20899, USA}\\

{\small \it ${}^2$Department of Physics, Georgetown University, Washington, DC 20057, USA}
\end{center}

\vspace{0.4cm}
\twocolumngrid

\textit{Time Dynamics.} In order to check that the DE accurately describes observables after relaxation, 
we compare the time dynamics of the central momentum peak, 
$\expv{\hat{n}_{k=0}(\tau) }=\bra{\psi(\tau)}\hat{n}_{k=0}\ket{\psi(\tau)}$, with the expectation value of 
$\hat{n}_{k=0}$ in the diagonal ensemble. Figure 1(a) shows that $\expv{\hat{n}_{k=0}(\tau) }$ relaxes to the diagonal 
prediction, with small fluctuations around this result, indicating that the diagonal ensemble correctly predicts 
the values of observables after relaxation. Additionally, the expectation value of $\hat{n}_{k=0}$ in the GME agrees 
with the DE, while the ME does not.  

\begin{figure}[!b]
\centerline{\includegraphics[scale=0.6]{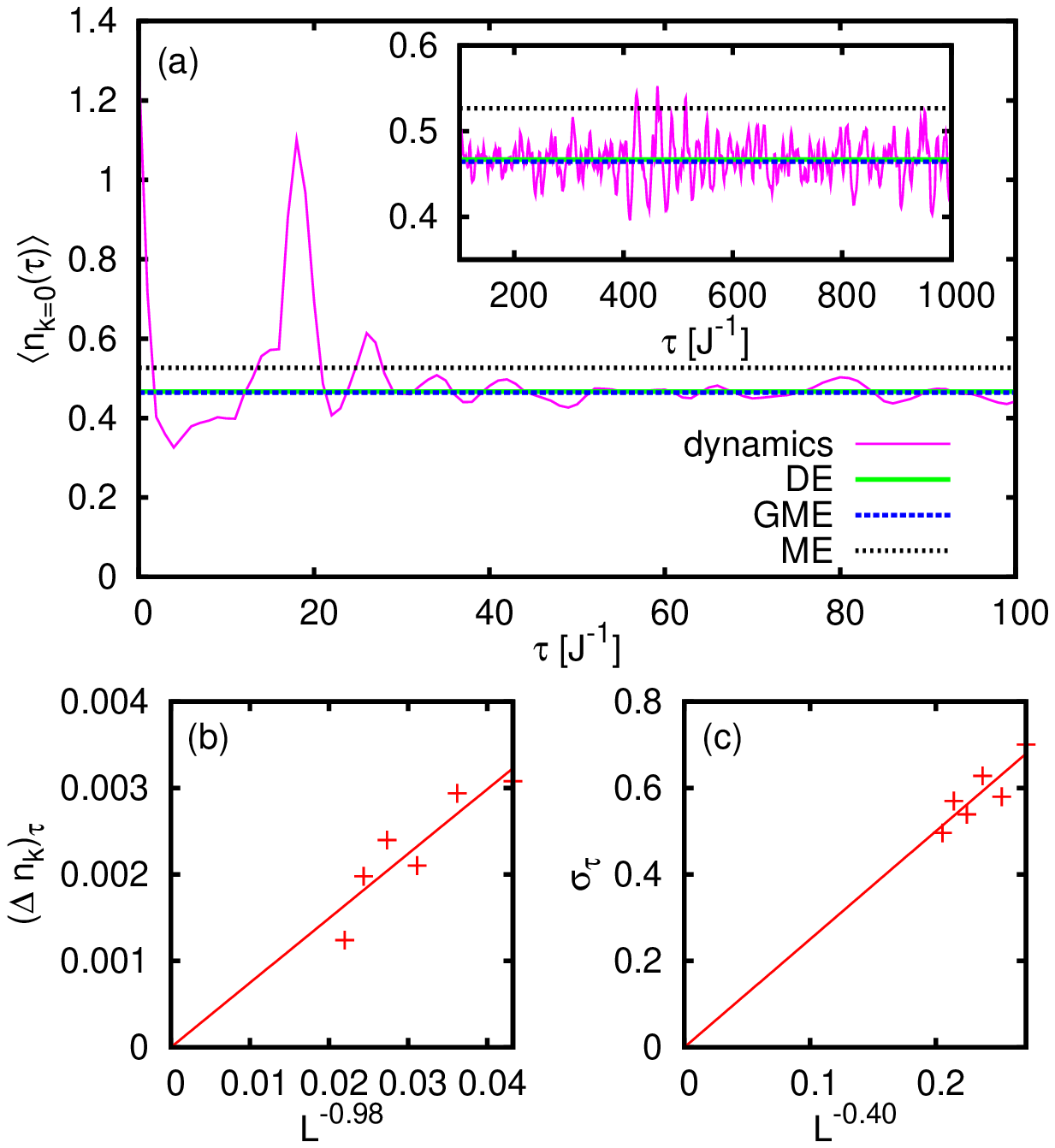}}
\caption{(a) Time evolution of $n_{k=0}$. Horizontal lines represent $\expv{\hat{n}_k}_\text{DE} $, 
$\expv{\hat{n}_k}_\text{GME} $, $\expv{\hat{n}_k}_\text{ME} $. $L=50$, $N=10$, $\varepsilon=0.72J$, 
$\delta_\text{GME}=0.8$, $\delta_\text{ME} = 0.05J$. Main Panel: $\tau \leq 100$. Inset: $100 \leq \tau \leq 1000$. 
(b) Integrated difference of the diagonal and time-averaged momentum distribution,$(\Delta n_k)_\tau$, 
versus $L^{-0.98}$, where a fit to $\sigma_\tau=zL^{-\gamma}$ gives $\gamma=0.98\pm0.25$. (c) Mean fluctuations 
of the  momentum distribution, $\sigma_\tau$, versus $L^{-0.4}$, where a fit to $(\Delta n_k)_\tau=zL^{-\gamma}$ 
gives $\gamma=0.40\pm0.12$. (b),(c) $\tau_1=100J, \tau_2=1000J, \varepsilon=0.72J$. $z$ is a generic multiplicative 
constant.}
\end{figure}

In order to understand the relation between the DE prediction and the actual time average over a finite time interval, 
$\overline{n_k}=\frac{1}{\tau_2-\tau_1}\int_{\tau_1}^{\tau_2} d\tau \expv{\hat{n}_k(\tau)}$, we study the integrated 
difference $(\Delta n_k)_\tau=\sum_k |\expv{\hat{n}_k}_\text{DE} - \overline{n_k}|/\sum_k \expv{\hat{n}_k}_\text{DE}$, 
and the time fluctuations $\sigma_{\tau}=\sum_k \sqrt{\overline{n_k^2}-\overline{n_k}^2}$ as a function of system size. 
$(\Delta n_k)_\tau$ is depicted in Fig.~1(b), where we also show the results of a fit to $(\Delta n_k)_\tau=zL^{-\gamma}$ 
with $\gamma=0.98\pm0.25$ ($z$ is a generic multiplicative constant). The integrated difference, $(\Delta n_k)_\tau$, 
clearly decreases with increasing system size and is several orders of magnitude smaller than similar comparisons with 
the statistical ensembles, confirming that degeneracies are irrelevant and the diagonal distribution accurately 
represents the long-time average.  Figure 1(c) depicts the time fluctuations of the momentum distribution, $\sigma_\tau$, 
where the exponent was determined from fitting $\sigma_\tau=zL^{-\gamma}$. This plots makes evident that the fluctuations 
also decrease with increasing system size (likely $\propto 1/\sqrt{L}$) and are expected to vanish in the thermodynamic 
limit.

\begin{figure}[!b]
\centerline{\includegraphics[scale=0.77]{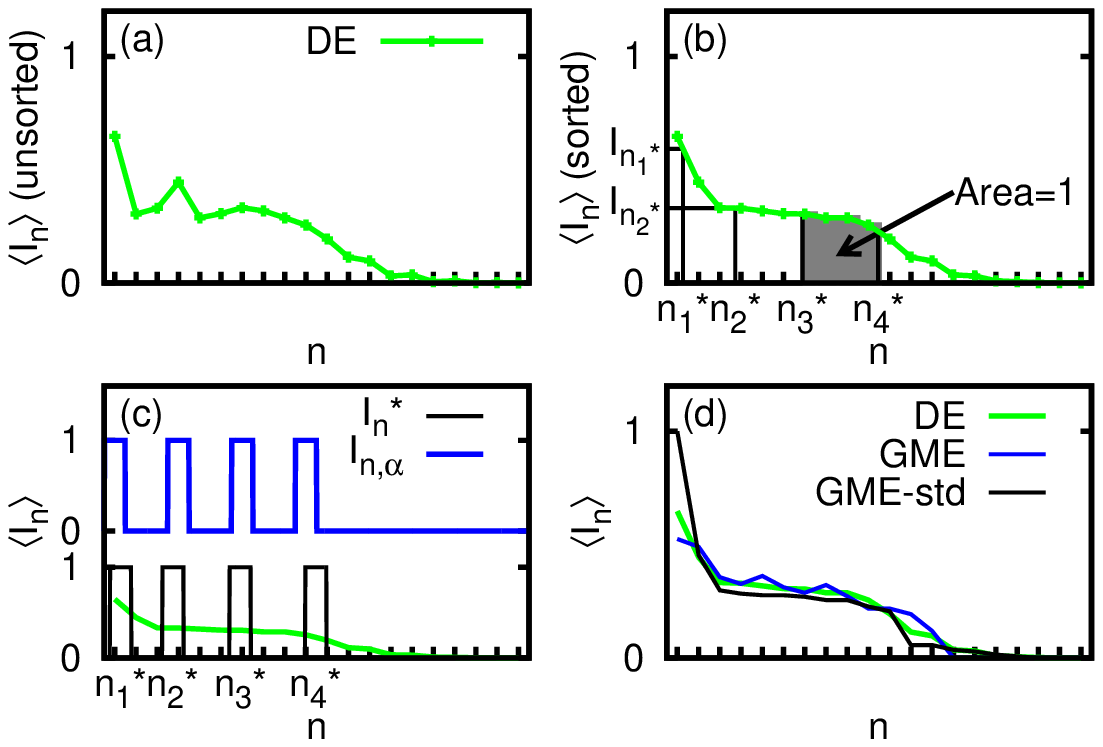}}
\caption{ (a) Distribution of conserved quantities in the diagonal ensemble, $\expv{I_n}_\textrm{DE}$, ordered by energy 
of the single-particle fermionic eigenstates. (b) $\expv{I_n}_\textrm{DE}$ sorted in descending order. The target distribution 
of $n_i^\ast$ are labeled as well as the corresponding $I_{n_i^\ast}$ for $i=1,2$. (c) Target distribution of conserved 
quantities, $I_n^\ast$, and distribution of eigenstate that minimized the distance $\delta_\alpha$ used to generate the 
GME, $I_{n,\alpha}$. (d) $\expv{I_n}$ in the DE and GME using method described in text as well as the standard metric 
(GME-std). $L=20, N=4, \varepsilon=0.72J$.}
\end{figure}

\textit{Constructing the generalized microcanonical ensemble.}
We expand upon our method for generating the generalized microcanonical ensemble with a series of plots in Fig.~2.  
Constructing the generalized microcanonical ensemble for a system of 1D hard-core bosons presents some unique challenges.  
In particular, the set of conserved quantities are the single-particle occupations of the underlying fermions used to 
build the many body eigenstates. Thus in any given eigenstate, each conserved quantity is either $0$ or $1$ corresponding to
whether or not that fermionic state is occupied, such as $I_{n,\alpha}$ in Fig.~2(c). On the other hand, the distribution 
in the diagonal ensemble is a continuous variable, $0 \leq \expv{I_n}_\textrm{DE}  \leq 1 $.  For each individual eigenstate, 
we must include or exclude it from the GME based on how close the discrete distribution in the eigenstate is to the 
continuous distribution of the quenched state. One possible method is a simple extension of the microcanonical metric 
\[\delta^\prime_\alpha =\left[  \sum_{n=1}^L (I_n^\ast - I_{n,\alpha})^2/\sum_n I_{n,\alpha}^2\right] ^{1/2}.\]  
We found this was not guaranteed to accurately describe the distribution of conserved quantities, particularly for 
low energies for the system sizes studied (see Fig. 2(d)). Instead of using the standard metric, we perform a weighted 
least-squares fit to a target distribution which is described as follows.

The unsorted conserved quantities of the quenched state $\expv{I_n}_\textrm{DE}$, ordered by energy of the single-particle 
fermionic eigenstates, are plotted in Fig.~2(a). As a first step, the distribution of $\expv{I_n}_\textrm{DE}$ is sorted 
in descending order as shown in Fig~2(b).  Next the values of $n_i^\ast$, which are not limited to integer values, are 
chosen so that $\int_{0.5}^{n_1^\ast} I(x)=0.5 $, where $I(x)=\expv{I_n}_\text{DE} $ for $ x  $ in the interval $( n-0.5, n+0.5]$.  
Subsequently the $n_i^\ast$ are determined so $\int_{n_{i-1}^\ast}^{n_i^\ast} I(x)=1 $ as depicted in Fig.~2(b). 
This set of $\{n_i^\ast\}$ determine the target distribution, labeled $I_n^\ast$ in Fig.~2(c). 
Each $n_i^\ast$ is assigned a weight, $I_{n_i^\ast}$, 
which is the interpolated value of $\expv{I_n}_\textrm{DE}$ at $n_i^\ast$.  The distance between the distribution of 
conserved quantities in eigenstate $\alpha$ and the distribution of the quenched state is then defined as
\begin{equation}\label{eq:gme}
\delta_\alpha=\left[\frac{1}{N}\sum_{i=1}^N I_{n^*_i}(n_{i,\alpha} - n^*_i)^2 \right] ^{1/2},
\end{equation}
where the $n_{i,\alpha}$ are the single particle eigenstates that are occupied in the many-body eigenstate $\alpha$.

Our method of constructing the GME is not unique.  We tested various different approaches, including an unweighted
sum in Eq.~\eqref{eq:gme}, etc, which gave similar results. We then employed the method which best 
reconstructed the distribution of conserved quantities over the full range of parameters studied. Note that for clarity Fig.~2 
displays data for the case of $N=4$ particles on $L=20$ sites.  There are only 4845 total eigenstates 
and 174 eigenstates used to construct the GME for the data shown. Even for such a small system, the 
distribution of conserved quantities within the GME is quite good. 

\paragraph{Lagrange multipliers and additivity.} In Fig.~3, we plot the expectation values of the conserved quantities 
in the diagonal ensemble along with corresponding Lagrange multipliers. The Lagrange multipliers are given by 
$\lambda_n=\ln\left(\frac{1-\expv{I_n}_\textrm{DE}}{\expv{I_n}_\textrm{DE}} \right)$. As can be seen, 
the Lagrange multipliers vary smoothly with the value of $I_n$ and with the index $n$ after the conserved 
quantities have been ordered in descending order. Additionally, the distribution of conserved quantities is very similar for 
two different lattice sizes with the same final energy-per-particle, when the index is normalized by the total number of 
conserved quantities. This is an important property of the Lagrange multipliers, which shows that even though the 
conserved quantities in these integrable systems are not additive in a strict sense, they can still be understood to be 
additive in a coarse grained sense, because the values of $\lambda_n$ are a smooth function of $I_n$. 

\begin{figure}[!h]
\centerline{\includegraphics[scale=0.6]{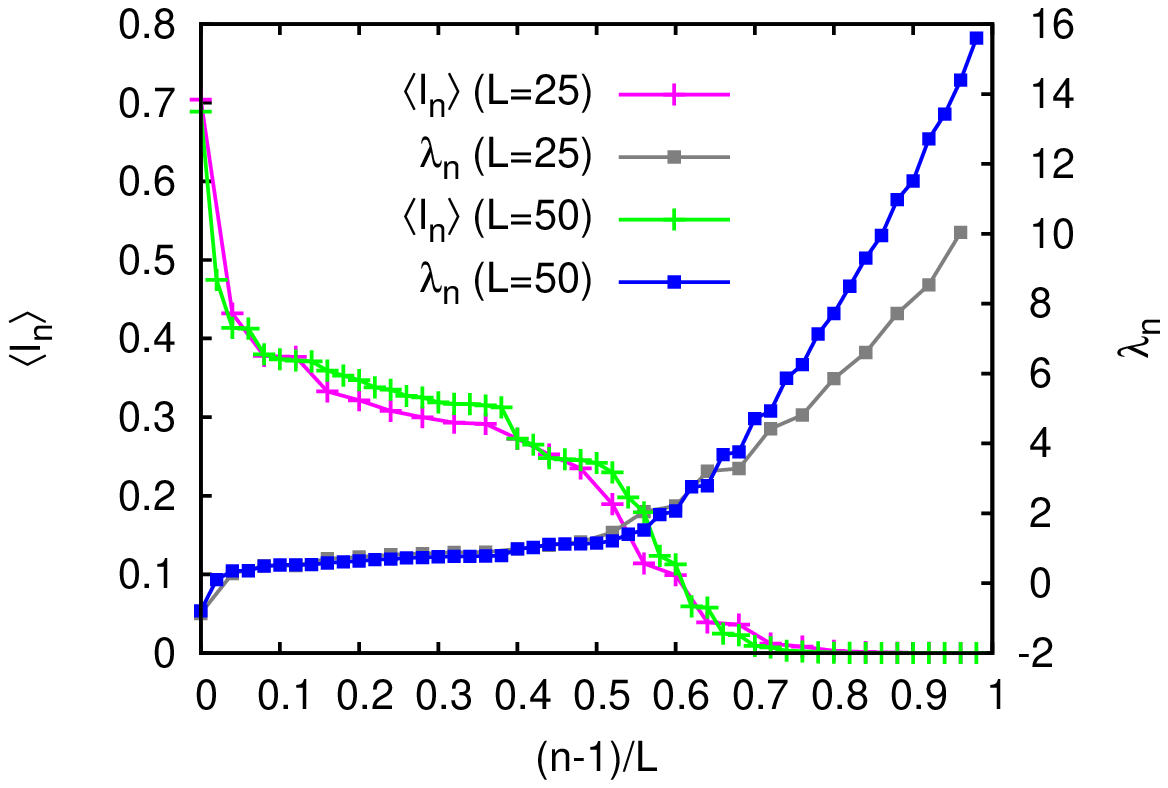}}
\caption{Expectation values of the conserved quantities in the diagonal ensemble and corresponding Lagrange 
multipliers for  $N=5$, $L=25$ and $N=10$, $L=50$. $\varepsilon=0.72t$.}
\end{figure}

\textit{Canonical Ensemble.} In addition to the results reported for the microcanonical distribution, we study the 
momentum distribution in the canonical ensemble (CE).  Within the canonical ensemble, observables are calculated as 
$\expv{\hat{A}}_\text{CE} = Z_\text{CE}^{-1}\text{Tr}\left[ \hat{A}e^{-\beta \hat{H}}  \right]$, where 
$Z_\text{CE}= \text{Tr}\left[e^{-\beta \hat{H} }  \right]$ and $\beta$ is the inverse temperature.  The temperature 
is calculated numerically so that $\expv{E}_\text{CE} = \bra{\psi_0}\hat{H}(\tau\geq 0)\ket{\psi_0} $.
In Fig.~4(a) we plot the integrated difference between the momentum distribution in the DE and CE, 
$(\Delta n_k)_{\text{CE}}$, for different lattice sizes as a function of the energy per particle of the quenched state. 
The comparison of the canonical and diagonal momentum distributions are similar to the comparison between the 
microcanonical and diagonal distributions, although there is some discrepancy between the microcanonical and 
canonical distributions due to finite size effects. In particular, the upturn in $(\Delta n_k)_\text{ME}$ at 
$\varepsilon= 1.4J$ is not present in $(\Delta n_k)_\text{CE}$.

\begin{figure}[!h]
\centerline{\includegraphics[scale=0.59]{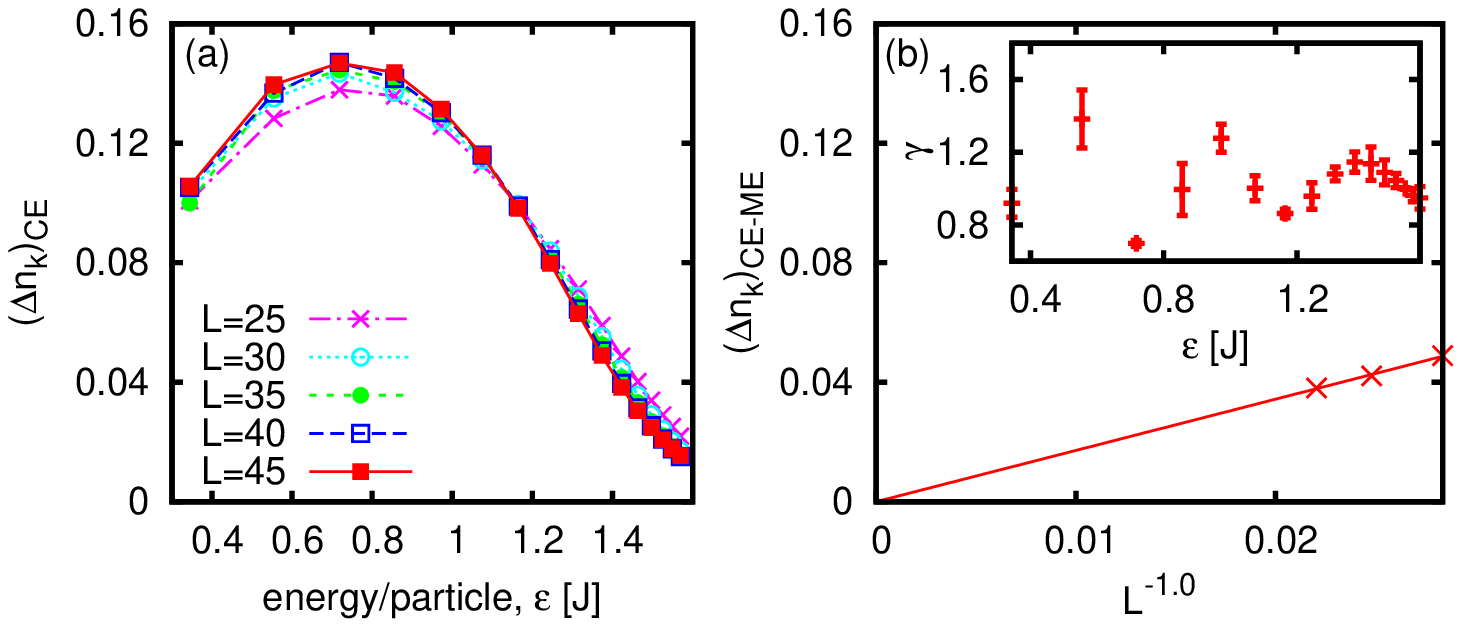}}
\caption{(a) Integrated difference between the diagonal and canonical momentum distributions, 
$\left( \Delta n_k\right)_\text{CE}$, versus energy per particle, $\varepsilon$. (b) Integrated difference 
of the momentum in the microcanonical and canonical ensembles, $\left( \Delta n_k\right)_\text{CE-ME}$, versus 
$L^{-1.19}$ for $\varepsilon=1.07J$, where a fit to $\left( \Delta n_k\right)_\text{CE-ME} = zL^{-\gamma}$ gives 
$ \gamma=1.0 \pm 0.07$. Inset: Scaling exponent, $\gamma$ of $\left( \Delta n_k\right)_\text{CE-ME}$ as a function 
of the energy per particle, $\varepsilon$.}
\end{figure}

In general, the microcanonical distribution provides us with a better description 
of the system because it is isolated, although it breaks down for very small systems because of poor statistics due 
to an insufficient number of states in the relevant energy window. Given the expectation that 
$\expv{\hat{n}_k}_\text{CE}$ and $\expv{\hat{n}_k}_\text{ME}$ are equal in the thermodynamic limit, we calculate the 
scaling exponent of the integrated difference of the momentum distribution,  
$\left( \Delta n_k\right)_\text{CE-ME} = zL^{-\gamma}$, where $ \left( \Delta n_k\right)_\text{CE-ME}=
\sum_k|\expv{\hat{n}_k}_\text{CE} - \expv{\hat{n}_k}_\text{ME}|/\sum_k\expv{\hat{n}_k}_\text{CE} $.  
In Fig.~4(b)  we plot $\left( \Delta n_k\right)_\text{CE-ME}$ versus $L^{-\gamma}$ for energy $\varepsilon=1.31J$, 
where $\gamma=1.0$.  In the inset, the exponent $\gamma$ is plotted as a function of the energy per particle. 
The mean value is $\gamma=1.03\pm 0.04$, which is consistent with $L^{-1}$ scaling for the difference between 
the canonical and grand canonical momentum distributions for hard-core bosons in a box found in Ref.~\cite{rigol05HCBc}.

On general grounds, the canonical and microcanonical ensembles are expected to equivalent in the thermodynamic limit, 
which is confirmed by our numerical results. We also see increasing agreement between the results of the generalized 
Gibbs and generalized microcanonical ensembles as system sizes increase.  There has been significant work in recent 
years on the equivalence of classical ensembles and less in quantum systems. In the classical case, 
Ellis \textit{et al.} have demonstrated that the microcanonical and canonical ensembles are equivalent if and only 
if the thermodynamic functions are equivalent, which is the case if the microcanonical entropy is concave 
\cite{ellis2000large}.  Furthermore, when the two ensemble are not equivalent, it may be possible to construct a 
generalized canonical ensemble, which contains an additional exponential which is a continuous function of the 
Hamiltonian, that is equivalent to the microcanonical ensemble \cite{costeniuc2005generalized}. 

\textit{Scaling of the GGE.} Given the trend observed in our results, and similar calculations in 
\cite{rigol07STATa} for much larger system sizes, we expect the integrated difference between the momentum 
distribution in the DE and GGE, $(\Delta n_k)_\text{GGE}$, to vanish in the thermodynamic limit. We fitted 
$(\Delta n_k)_\text{GGE}=zL^{-\gamma}$ and report the results for $\varepsilon=1.13J$  in the main text.  
In Fig.~5 we plot the exponent, $\gamma$, as a function of the energy per particle of the quenched state for 
all energies studied. The mean value of the exponent is $\gamma=0.731\pm 0.007$.

\begin{figure}[!h]
\centerline{\includegraphics[scale=0.65]{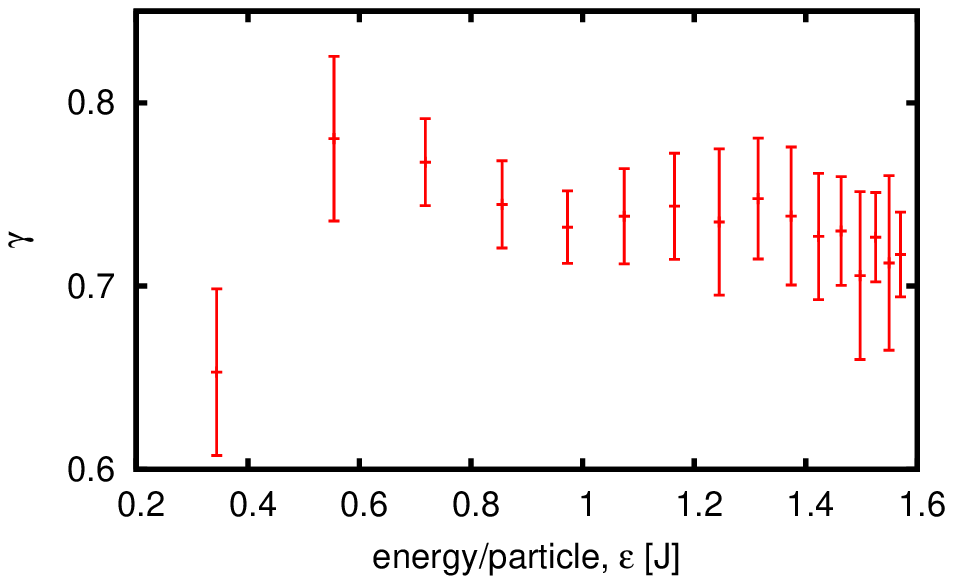}}
\caption{Exponent of power-law scaling of $(\Delta n_k)_\text{GGE}$ versus inverse system size as a function of the 
energy per particle.}
\end{figure}

\textit{Number of States.} The total number of states in our system, the number of states in the ME, and the 
number of states in the GME all scale exponentially with the system size as $z^L$.  For the total number of states, 
$L!/\left[N!(L-N)!\right]$, using Stirling's approximation, $z\approx (1-\nu)^{\nu-1}\nu^{-\nu}\approx 1.65$, where 
$\nu$ is the filling factor.  We confirm this scaling by numerically fitting to the total number of states.  
For $\varepsilon=0.72J$, we find numerically that $z\approx1.51$ in the ME, ($\delta_{ME}=0.05J$) and 
$ z\approx 1.37$ in the GME ($\delta_{GME}=0.8$).  Thus the number of states in the ME and GME windows as a the 
fraction of the total number of states vanishes in the thermodynamic limit.  Additionally the ratio of states 
in GME to ME vanishes.  These values are typical, although the precise exponents depend on the choice of the ME 
and GME window sizes.

\begin{figure}[!b]
\centerline{\includegraphics[scale=0.6]{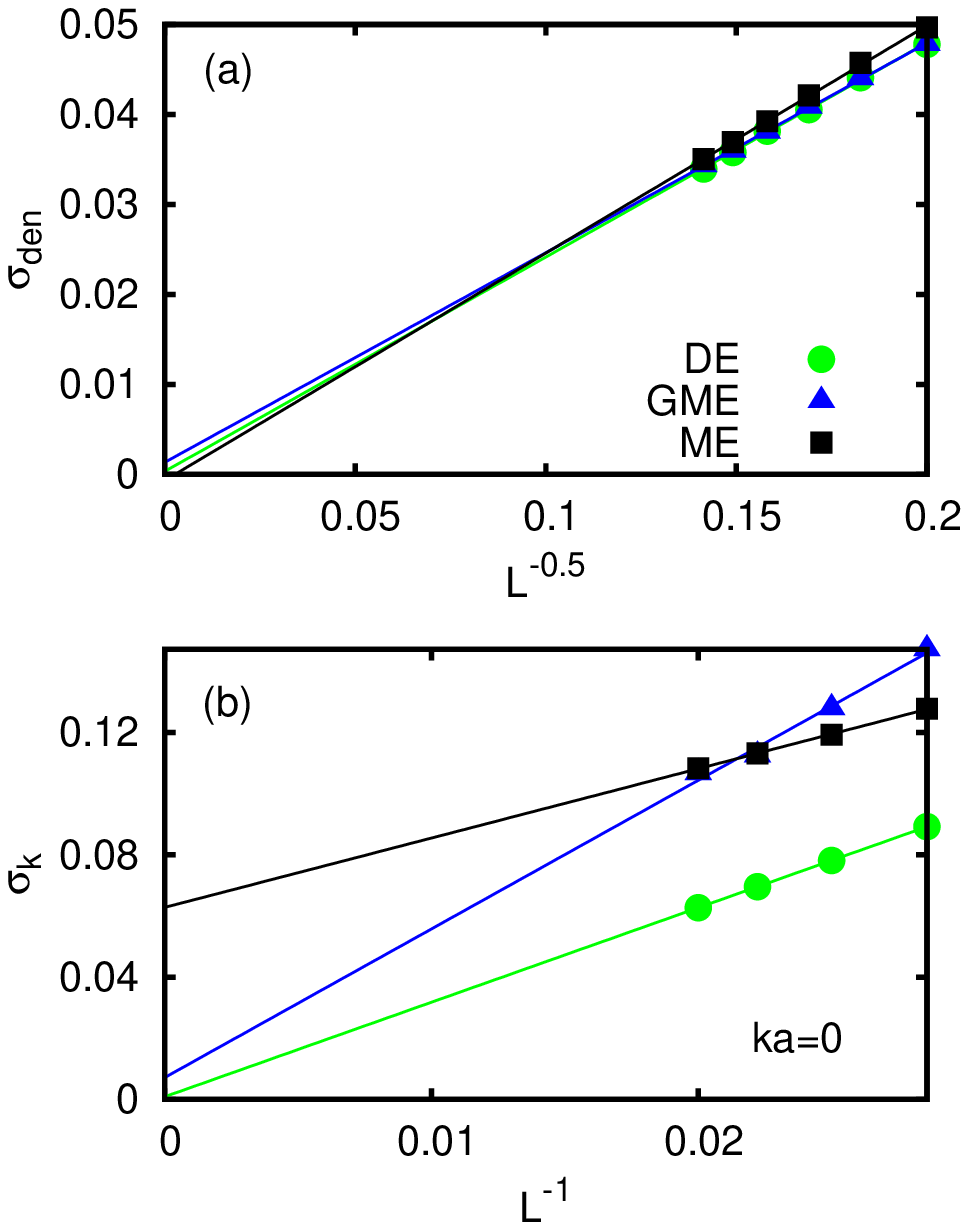}}
\caption{(a) Average density fluctuations, $\sigma_{\text{den}}$ versus $L^{-0.5}$ for fixed energy per particle 
in the diagonal (DE), generalized microcanonical (GME) and microcanonical (ME) ensembles. (b)  Momentum fluctuations, 
$\sigma_k$ for $ka=0$. $\varepsilon= 0.72J$, $\delta_\text{ME}=0.05J, \delta_\text{GME}=0.8.$}
\end{figure}

\textit{Fluctuations of Local Observables.} For local observables, the fluctuation of the eigenstate to eigenstate 
expectation values are expected to scale as $L^{-1/2}$ \cite{biroli2009}.  In Fig.~6(a) we plot the fluctuations of the
site occupations averaged over five sites, 
$\sigma_{\text{den}}=1/5 \sum_{i=1}^{5}\sqrt{\expv{\hat{n}_{x=iN}^2} - \expv{\hat{n}_{x=iN}}^2}$ 
vs. $L^{-0.5}$ in the DE, GME, and ME distributions along with a linear fit to $\sigma_\text{den}$ versus $L^{-0.5}$.  
The data strongly suggests that the fluctuations scale as $L^{-0.5}$ as predicted in Ref.~\cite{biroli2009} and will 
vanish in the thermodynamic limit for all three ensembles.

In the main text, we presented evidence that the fluctuations of the eigenstate to eigenstate expectation value of
the momenta occupations vanish in the thermodynamic limit for the DE and GME.  We have also studied how the width of the 
distribution of the momenta eigenstate expectation values, $\sigma_k$, scales for the ME. In Fig.~6(b), we plot the 
fluctuations of $\hat{n}_{ka=0}$.  For the microcanonical ensemble it is difficult to reach a conclusion as to whether 
it vanishes or remains finite in the thermodynamic limit. We do, however, find that for this observable the results 
are clearly different from those for observables that only contain short range correlations, and that the behavior 
of the width of its distribution within the ME behaves quite differently from the one within the
DE and the GME. In the latter two the width decreases much rapidly with system size.


\begin{thebibliography}{10}
\bibitem{In_integer}In each many-body eigenstate of the Hamiltonian, $\ket{\alpha}$, $I_{n,\alpha}=\bra{\alpha}\hat{I}_n\ket{\alpha}$ can only be zero or one, because they are constructed by occupying single-particle fermionic states.
\bibitem{nstar}We compute the $n^*_i$, which are not restricted to integers, as follows. The first one ($n^*_1$) is computed so that $\int_{0.5}^{n^*_1}I(x)dx=0.5$,  where $I(x)=I_n, x \in ( n-0.5, n+0.5]$. For all other values of $n^*_i$, 
 $\int_{n^*_{i-1}}^{n^*_{i}}I(x)dx=1$.
\end{thebibliography}

\end{document}